\documentclass[useAMS,usenatbib]{mn2e}
\usepackage{amssymb}
\title[Dynamics of Hot Accretion Flow]{Dynamics of Hot Accretion Flow 
with Thermal Conduction}
\author[K. Faghei]{Kazem Faghei\thanks{E-mail:kfaghei@du.ac.ir}\\
School of Physics, Damghan University, Damghan, Iran\\
}
\begin{document}


\pubyear{2011}

\maketitle

\label{firstpage}

\begin{abstract}
The purpose of this paper is to explore the dynamical behaviour of hot accretion flow with thermal conduction.
The importance of thermal conduction on hot accretion flow is confirmed by observations of the hot gas that surrounds Sgr A$^*$
and a few other nearby galactic nuclei. In this research,
the effect of thermal conduction is studied by a saturated form of 
it, as is appropriate for weakly collisional systems.
The angular
momentum transport is assumed to be a result of viscous turbulence and the $\alpha$-prescription is used for
the kinematic coefficient of viscosity.
The equations of accretion flow are solved in a simplified
one-dimensional model that neglects the latitudinal dependence of the flow. To solve the
integrated equations that govern the dynamical behaviour of the accretion flow, we have used an
unsteady self-similar solution. The solution provides some insights into the dynamics of quasi-spherical
accretion flow and avoids from limits of the steady self-similar solution.
In comparison to accretion flows without thermal conduction, 
the disc generally becomes cooler and denser. These properties are 
qualitatively consistent with performed simulations in hot accretion flows. 
Moreover, the angular velocity increases with the magnitude of conduction, while the radial infall velocity decreases.  
The mass accretion
rate onto the central object is reduced in the presence of thermal conduction. 
We found that the viscosity and thermal conduction have the opposite effects on the physical variables. 
Furthermore, the flow represents a transonic point that moves inward 
with the magnitude of conduction or viscosity.
\end{abstract}


\section{Introduction}
Accretion is one of the most important physical processes in astrophysics. It is widely accepted that the accreting matter 
toward the central object is a source of power active galactic nuclei (AGN), and galactic X-ray sources 
(see for a review e.g. Frank et al. 2002). This idea is also well-applicable to interpret many observations of astrophysical phenomena, 
such as, prototype stellar objects (Chang \& Choi 2002), symbiotic stars (Lee \& Park 1999), gamma-ray bursts (Brown et al. 2000). There
are some types of accretion flows that a significant fraction of the generated heat  by dissipation processes retains in the fluid rather 
than being radiated away, have been the subject of considerable attention in recent years (Ogilvie 1999; 
Chang 2005, Akizuki \& Fukue 2006, Khesali \& Faghei 2009, Faghei 2011).
These advection dominated accretion flows (ADAF) place an intermediate position between the spherically symmetric accretion flow of
 non-rotating fluid (Bondi 1952) and the cool thin disc of classical accretion disc theory (e. g. Pringle 1981). These types of accretion flows
have been widely applied to explain observations of the galactic black hole candidate (e. g. Narayan et al. 1996, Hameury et al. 1997),
the spectral transition of Cyg x-1 (Esin 1996) and multi-wavelength spectral properties of Sgr A$^*$ (Narayan \& Yi 1995; Manmoto et al. 2000;
Narayan et al. 1997).

The X-ray observations of black holes imply that they are capable of accreting gas under a variety of flow configurations. In particular,
observational evidences confirm existence of hot accretion flow, contrasted with the classical cold and thin accretion disc scenario
(Shakura \& Sunyaev 1973). Hot accretion flow can be found in the population of supermassive black holes in galactic nuclei and during 
quiescent of accretion onto stellar-mass black holes in X-ray transients (e.g.,
Narayan et al. 1998a; Lasota et al. 1996; Di Matteo et al. 2000;
Esin et al. 1997, 2001; Menou et al. 1999; see Narayan et al.
1998b; Melia \& Falcke 2001; Narayan 2002; Narayan \& Quataert 2005 for reviews). \textit{Chandra} observations provide tight
constraints on the density and temperature of gas at or near the Bondi capture radius in Sgr A$^*$ and several other nearby galactic nuclei. 
Tanaka \& Menou (2006) used these constraints (Loewenstein et al. 2001; Baganoff
et al. 2003; Di Matteo et al. 2003; Ho et al. 2003) to calculate the mean free path for the observed gas. They suggested that accretion
in these systems will be proceeded under the weakly collisional condition. Furthermore, they suggested that thermal conduction can be as
a possible mechanism by which the sufficient extra heating is provided in hot advection dominated accretion flows.

Generally, semi-analytical 
studies of hot accretion flows with thermal conduction have been related to steady state models
(e. g. Tanaka \& Menou 2006; Johnson \& Quataert 2007; Shadmehri 2008; Abbassi et al. 2008, 2010; Ghanbari et al. 2009), and
dynamics of such systems have been studied in simulation models (e. g. Sharma et al. 2008; Wu et al. 2010). For example,  
Tanaka \& Menou (2006) have carried out a related analysis and
found the accretion flow can spontaneously produce thermal
outflows driven in part by conduction. Their analysis is two-dimensional but self-similar
in radius. Their assumption of self-similarity enforces a density profile that varies as $r^{-3/2}$,
whereas simulations of ADAFs consistently find density profiles
shallower than this (e.g., Stone et al. 1999; Igumenshchev \&
Abramowicz 1999; Stone \& Pringle 2001; Hawley \& Balbus 2002;
Igumenshchev et al. 2003). Johnson \& Quataert (2007) studied the effects of electron thermal conduction 
on the properties of hot accretion flows, under the assumption of spherical symmetry. 
Since, electron heat conduction is important for low accretion rate systems, thus their model
is applicable for Sgr A$^*$ in the Galactic centre.
They show that heat conduction leads to supervirial
temperatures, implying that conduction significantly modifies the structure of the accretion flow.
Their model similar to Tanaka \& Menou (2006) was the steady state, but they solved their equations numerically.

As mentioned, semi-analytical studies of hot accretion flows with thermal conduction have been  
in a steady state. Thus, 
it will be interesting to study \textit{dynamics} of such systems. 
Ogilvie (1999) by the unsteady self-similar method studied time-dependence of quasi-spherical accretion
flow without thermal conduction.
The solutions of Ogilvie (1999) provided some insight into the dynamics of quasi-spherical accretion and
 avoided many of the limits of the steady self-similar
solution. In this research, we want to explore how thermal conduction can affect the dynamics of a rotating
and accreting viscous gas. We
answer this question by solving Ogilvie (1999) model that is affected by thermal conduction. 
This paper is organized as
follows. In Section 2, we define the general problem of constructing
a model for hot quasi-spherical accretion flow.
In Section 3, we use the unsteady self-similar method to solve the integrated
equations that govern the dynamical behaviour of the accreting gas, and numerical study of the model is brought 
in this section, too.
We will present a summary of the model in Section 4.

\section{Basic Equations}
We start with the approach adopted by Ogilvie (1999), who studied
quasi-spherical accretion flows without thermal conduction. 
Thus, we derive the basic equations that describe the
physics of accretion flow with thermal conduction. We use
the spherical coordinates $(r, \theta, \phi)$ centred on the accreting object
and make the following standard assumptions:

\begin{enumerate}
 \item The gravitational force on a fluid element is characterized
by the Newtonian potential of a point mass, $\Psi=-G M_* / r$, with
$G$ representing the gravitational constant and $M_*$ standing for the
mass of the central star.

\item The written equations  in spherical coordinates are 
considered in the equatorial plane $\theta=\pi/2$ and terms with any $\theta$ and $\phi$
dependence are neglected, hence all quantities will be expressed in
terms of spherical radius $r$ and time $t$.

\item For simplicity, self-gravity and general relativistic effects
have been neglected.
\end{enumerate}

Under these assumptions,
the dynamics of accretion flow describes by the following equations:
\\
the continuity equation
\begin{equation}\label{a1}
\frac{\partial\rho}{\partial t}
+\frac{1}{r^{2}}\frac{\partial}{\partial r}(r^{2}\rho v_{r})=0,
\end{equation}
the radial force equation
\begin{equation}\label{a2}
\frac{\partial v_{r}}{\partial t}+v_{r}\frac{\partial
v_{r}}{\partial r}=r(\Omega^{2}-\Omega_K^{2})-\frac{1}{\rho}\frac{\partial p}{\partial r},
\end{equation}
the azimuthal force equation
\begin{equation}\label{a3}
\rho\left[\frac{\partial}{\partial
t}(r^{2}\Omega)+v_{r}\frac{\partial}{\partial r}(r^{2}\Omega)\right]=
  \frac{1}{r^{2}}\frac{\partial}{\partial r}\left[\nu\rho r^{4}\frac{\partial \Omega}{\partial r}\right],
\end{equation}
the energy equation
\begin{eqnarray}
 \nonumber \frac{1}{\gamma-1}\left[\frac{\partial p}{\partial t}+v_r\frac{\partial p}{\partial r}\right]+
\frac{\gamma}{\gamma-1}\frac{p}{r^2}\frac{\partial}{\partial r}\left(r^2 v_r\right)=\\ 
Q_{vis}-Q_{rad}+Q_{cond}.
\end{eqnarray}
Here $\rho$ the density, $v_r$ the radial velocity, $\Omega$ the angular velocity, $\Omega_K[=(G M_*/r^3)^{1/2}]$ 
Keplerian angular velocity, $p$ the gas pressure, $\gamma$ is the adiabatic index,
$\nu$ the kinematic viscosity coefficient and it is given as in Narayan \& Yi (1995a) by an $\alpha$-model 

\begin{equation}
   \nu = \alpha \frac{p_{gas}}{\rho\Omega_{K}}.
\end{equation}
The parameter of $\alpha$ is assumed to be a constant less than unity. 
The terms on the right-hand side of the energy equation, $Q_{vis}$ is the heating rate of 
 the gas by the viscous dissipation, $Q_{rad}$ represents the energy loss through radiative
cooling, and $Q_{cond}$ is the transported energy  by thermal conduction.
For the right-hand side of the energy equation, we can write
 \begin{equation}
   Q_{adv}=Q_{vis}-Q_{rad}+Q_{cond}
\end{equation}
where $Q_{adv}$ is the advective transport of energy. We employ the advection factor, $f=1-Q_{rad}/Q_{vis}$, 
that describes the fraction of the dissipation energy which is stored in the accretion flow and advected into the central
object rather than being radiated away. The advection factor of $f$ in general depends on
the details of the heating and radiative cooling mechanism and will vary with position (e.g.
Watari 2006, 2007; Sinha et al. 2009). However, we assume a constant $f$ for simplicity. 
Clearly, the case $f = 1$ corresponds to the extreme limit of no radiative
cooling and in the limit of efficient radiative cooling, we have $f = 0$. 

In a collisional plasma, mean free path for electron energy exchange, $\lambda$, is shorter than temperature scale 
height, $L_T=T/|\nabla T|$, and thus the heat flux due to thermal conduction can be written as
  \begin{equation}
   F_{cond}=-\kappa \nabla T,
\end{equation}
where $\kappa$ is the thermal conductivity coefficient. Thermal conductivity in a dense, fully ionized gas is given by the Spitzer (1962)
formula,
  \begin{equation}
   \kappa=\frac{1.84 \times 10^{-5} T_e^{5/2}}{\ln \Lambda},
\end{equation}
where $T_e$ is the electron temperature ($T_e=T$ for a one-temperature plasma) and $\ln \Lambda$ is Coulomb logarithm that for
 $T > 4.2 \times 10^5 K $ is
  \begin{equation}
   \ln \Lambda=29.7+\ln n^{-1/2} (T_e/10^6 K).
\end{equation}
The heat is conducted by the electron, and equation (8) includes the effect of the self-consistent electric required to maintain the electric
current at zero; this reduces $\kappa$ by a factor of about $0.4$ from the value it would otherwise have (Cowie \& McKee 1977, hereafter CM77).

As noted in the introduction, the inner regions of hot accretion flows are, in many cases, collisionless with electron 
mean free path due to Coulomb collision larger than the radius (e. g. Tanaka \& Menou 2006). 
When the mean free path of an electron 
becomes comparable to or larger than the temperature gradient scale $\lambda \gtrsim T/|\nabla T|$, equation (7) for the heat flux
is no longer valid; CM77 described this effect as \textit{saturation}.  The maximum heat flux in a plasma can be expressed as 
$(3/2) n_e k T_e v_{char}$, where $v_{char}$ is a characteristic velocity which one might expect to be the order of 
the electron thermal velocity (Parker 1963). Assuming Maxwellian distribution for heat source, the characteristic velocity can be written
as (Williams 1971; CM77)
   \begin{equation}
   V_{char}=(\frac{8}{9\,\pi})^{1/2}\,(\frac{k\,T}{m_e})^{1/2}.
\end{equation}
Similar to CM77, we assume that the heat flux is reduced by the same factor of $0.4$ in the saturated case as in the classical (collisional)
case so that the saturated heat flux is 
\begin{equation}
   F_{sat}=0.4\,n_e k T_e\,\sqrt{\frac{2 k T_e}{\pi m_e}}.
\end{equation}
CM77 showed that the saturated heat flux is significantly less than conjectured by Parker (1963). Thus, in order to explicitly
allow for uncertainty in the estimate of $F_{sat}$, they introduced a factor of $\phi_s$, which was less than unity and rewrote equation (11)
as 
\begin{equation}
   F_{sat}=5 \phi_s \rho c_s^3=5 \phi_s p \sqrt{\frac{p}{\rho}},
\end{equation}
where $c_s$ is sound speed, which is defined as $c_s^2=p / \rho$.
The factor of $\phi_s$ is called as saturation constant (CM77).
Now,    
the viscous heating rate  
and the energy transport by thermal conduction 
are expressed as
\begin{equation}
   Q_{vis}=\nu\rho r^2 \left(\frac{\partial \Omega}{\partial  r}\right)^2
\end{equation}
\begin{equation}
   Q_{cond}=-\frac{1}{r^2}\frac{\partial}{\partial r} \left(r^2 F_{sat}\right)
\end{equation}
By using equations (13) and (14) for the advective transport of energy,
we can write
 \begin{equation}
   Q_{adv}=f\nu\rho r^2 \left(\frac{\partial \Omega}{\partial  r}\right)^2-\frac{1}{r^2}\frac{\partial}{\partial r} \left(r^2 F_{sat}\right)
\end{equation} 

The mass accretion rate in a qausi-spherical accretion flow can be written as
 \begin{equation}
   \dot{M}(r,t)=-4\pi r^2 \rho v_r.
\end{equation}
We will use this quantity in the next section, and will investigate effects of saturation constant and 
viscous parameter on it.

\section{Self-Similar Solutions}
\subsection{analysis}
Tanaka \& Menou (2006) solved essentially equations (1)-(4) for the case of a steady, radially self-similar flow. Here, we will try to find
unsteady self-similar solutions for these equations. Thus, we introduce a similarity variable $\xi$ and assume that each physical quantity
is given by the following form:
 \begin{equation}
  \xi=r(G M_* t^2)^{-1/3},
\end{equation} 
 \begin{equation}
  \rho(r,t)=R(\xi) (\dot{M}_0 / G M_*) t^{-1},
\end{equation} 
\begin{equation}
  p(r,t)=\Pi(\xi) (\dot{M}_0 / (G M_*)^{1/3}) t^{-5/3},
\end{equation}
\begin{equation}
  v_r(r,t)=V(\xi) (G M_*)^{1/3} t^{-1/3},
\end{equation}
\begin{equation}
 \Omega(r,t)=\omega(\xi) t^{-1},
\end{equation}
\begin{equation}
 \dot{M}(r,t)=\dot{M}_0\dot{m}(\xi),
\end{equation}
where $\dot{M}_0$ is a constant and its value can be obtained by typical values of the system. In addition, we assumed that $\dot{M}(r,t)$
under similarity transformations is a function of $\xi$ only (Khesali \& Faghei 2008, 2009). Substitution of above transformations
into the basic equations (1)-(4), yields  dimensionless equations below,
\begin{equation}\label{a20}
\left(V-\frac{2\xi}{3}\right)\frac{dR}{d\xi}-R=-\frac{R}{\xi^2}
\frac{d}{d\xi}\left(\xi^2V\right),
\end{equation}
\begin{eqnarray}\label{a21}
\left(V-\frac{2\xi}{3}\right)\frac{dV}{d\xi}-\frac{V}{3}=
\xi(\omega^{2}-\xi^{-3})-\frac{1}{R}\frac{d\Pi}{d\xi},
\end{eqnarray}
\begin{eqnarray}\label{a22}
\nonumber R\left[\left(V-\frac{2\xi}{3}
\right)\frac{d}{d\xi}\left(\xi^{2}\omega\right)
+\frac{1}{3}\left(\xi^{2}\omega\right)\right]~~~~~~~~~~~~~~~~~~~~~\\
=\frac{\alpha}{\xi^{2}}\frac{d}{d\xi}
\left[\Pi\xi^{11/2}\frac{d\omega}{d\xi}\right],
\end{eqnarray}

\begin{eqnarray}
\nonumber\frac{1}{\gamma-1}\left[\left(V-\frac{2\xi}{3}\right)\frac{d\Pi}{d\xi}-\frac{5}{3}\Pi\right]+
\frac{\gamma}{\gamma-1}\frac{\Pi}{\xi^2}\frac{d}{d\xi}\left(\xi^2V
\right)\\
= \alpha f \Pi \xi^{7/2}\left(\frac{d\omega}{d\xi}\right)^2
-\frac{5 \phi_s}{\xi^2} \frac{d}{d \xi} \left(\xi^2 \Pi \sqrt{\frac{\Pi}{R}} \right).
\end{eqnarray} 
These equations provide a fourth-order system of non-linear ordinary differential equations that must be 
solved numerically.

\subsection{Inner limit}
An appropriate asymptotic solution as $\xi\rightarrow0$  is the form as
\begin{equation}
    R(\xi) \sim \xi^{-3/2} (R_{0}+R_{1} \xi + \cdot \cdot \cdot),
\end{equation}
\begin{equation}
    \Pi(\xi) \sim \xi^{-5/2} (\Pi_{0}+\Pi_{1} \xi + \cdot \cdot \cdot),
\end{equation}
\begin{equation}
    V(\xi) \sim \xi^{-1/2} (V_{0}+V_{1} \xi + \cdot \cdot \cdot),
\end{equation}
\begin{equation}
    \omega(\xi) \sim \xi^{-3/2} (\omega_{0}+\omega_{1} \xi + \cdot \cdot \cdot),
\end{equation}
in which undetermined coefficients of $R_0$, $R_1$, $\Pi_0$, and etc must be specified. 
By substituting above relations in equations (23)-(26) and choosing the significant sentences, we can write
\begin{equation}
    V_0^2+\frac{5 \Pi_0}{R_0}-2+2\,\omega_0^2\approx 0,
\end{equation} 
\begin{equation}
    2\,R_0\,V_0 +3\,\alpha\,\Pi_0\approx0,
\end{equation}
\begin{equation}
    6\,V_0\,(\gamma-\frac{5}{3})+(\gamma-1)\left[20\phi_s \sqrt{\frac{\Pi_0}{R_0}}-9 \alpha f \omega_0^2 \right]\approx0.
\end{equation}
Also, the dimensionless mass accretion rate, $\dot{m}(\xi)$, under the above asymptotic solution becomes
\begin{equation}
    \dot{m}_{in}\approx-4 \pi R_0 V_0,
\end{equation}
where $\dot{m}_{in}$ is the value of $\dot{m}$ at $\xi_{in}$, where $\xi_{in}$ is a point near to the centre.
After algebraic manipulations for equations (31)-(34), we obtain an algebraic equation for $R_0$:
\begin{eqnarray}
 \nonumber R_0^2
-\frac{10}{27} \frac{\phi_s}{\alpha f} \sqrt{\frac{6 \dot{m}_{in}}{\alpha \pi}}\, R_0^{3/2}
-\frac{5}{12}
\frac{\dot{m}_{in}}{\alpha \pi}  \times \\
\left( 1-\frac{2}{5 f} \frac{\gamma-5/3}{\gamma-1} \right) R_0 
-\frac{1}{32}
\left(\frac{\dot{m}_{in}}{\pi}\right)^2\approx0,
\end{eqnarray}
and the rest of the physical variables are
\begin{equation}
    \Pi_0\approx\frac{\dot{m}_{in}}{6\pi\alpha},
\end{equation}
\begin{equation}
   V_{{0}}\approx-\frac{\dot{m}_{in}}{4\pi R_0},
\end{equation}
\begin{equation}
    \omega_0^2\approx1-\frac{5}{12} \frac{\dot{m}_{in}}{\pi\alpha R_0} \left( 1+\frac{3}{40} \frac{\alpha \dot{m}_{in}}{\pi R_0} \right).
\end{equation}
Without thermal conduction, $\phi_s=0$, equation (35) can be solved analytically.
Since, we want to consider systems with non-zero saturation constant, $\phi_s\neq 0$, we will solve this equation numerically.

\input{epsf}
\begin{figure*}
\centerline
{ 
{\epsfxsize=7.5cm\epsffile{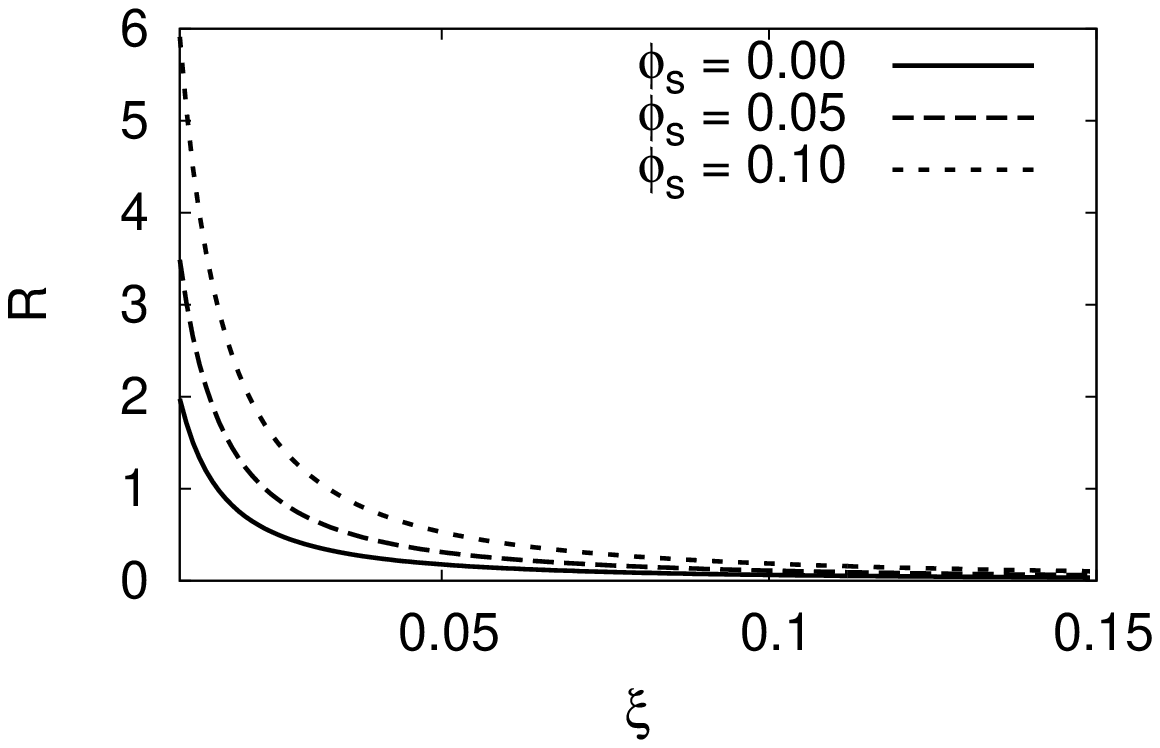}  }{\epsfxsize=7.5cm\epsffile{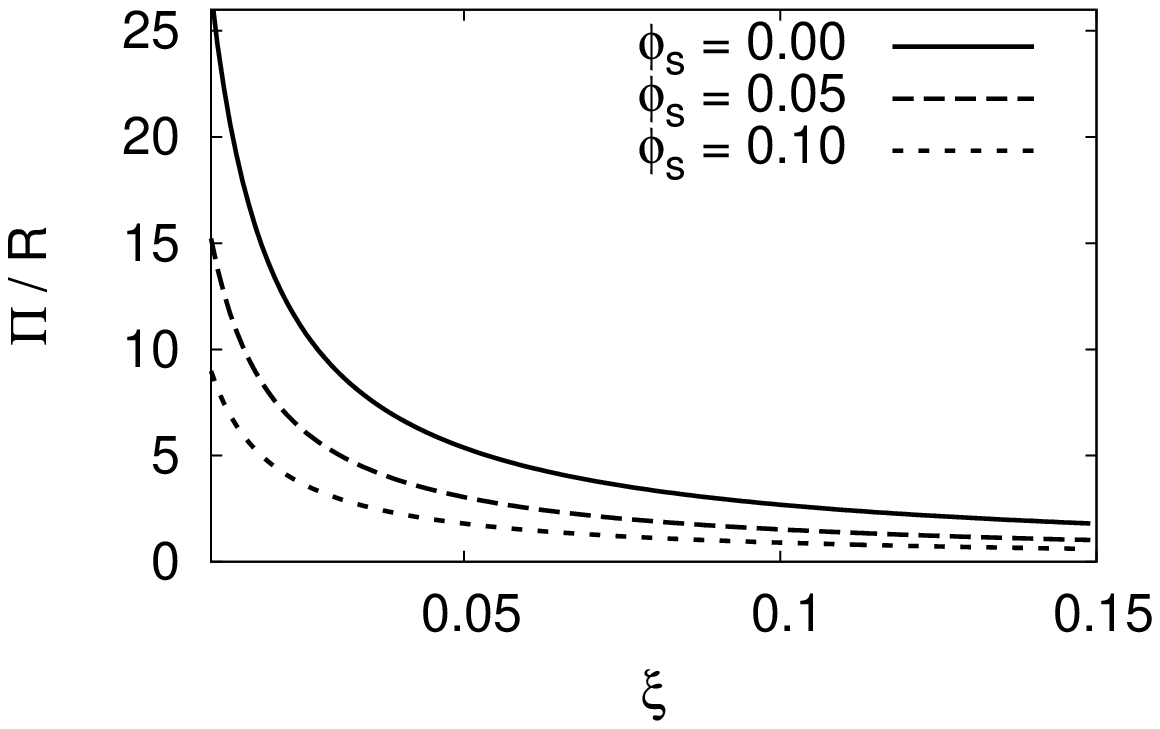}  }
} 
\centerline
{ 
{\epsfxsize=7.5cm\epsffile{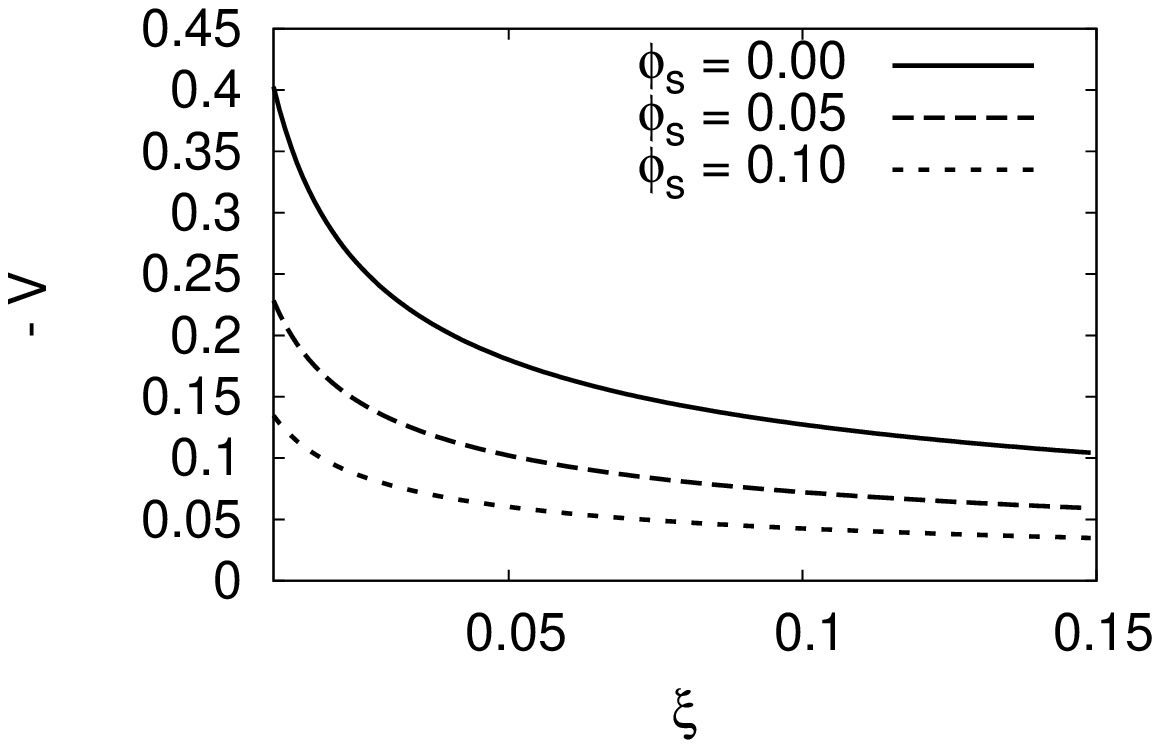}  }{\epsfxsize=7.5cm\epsffile{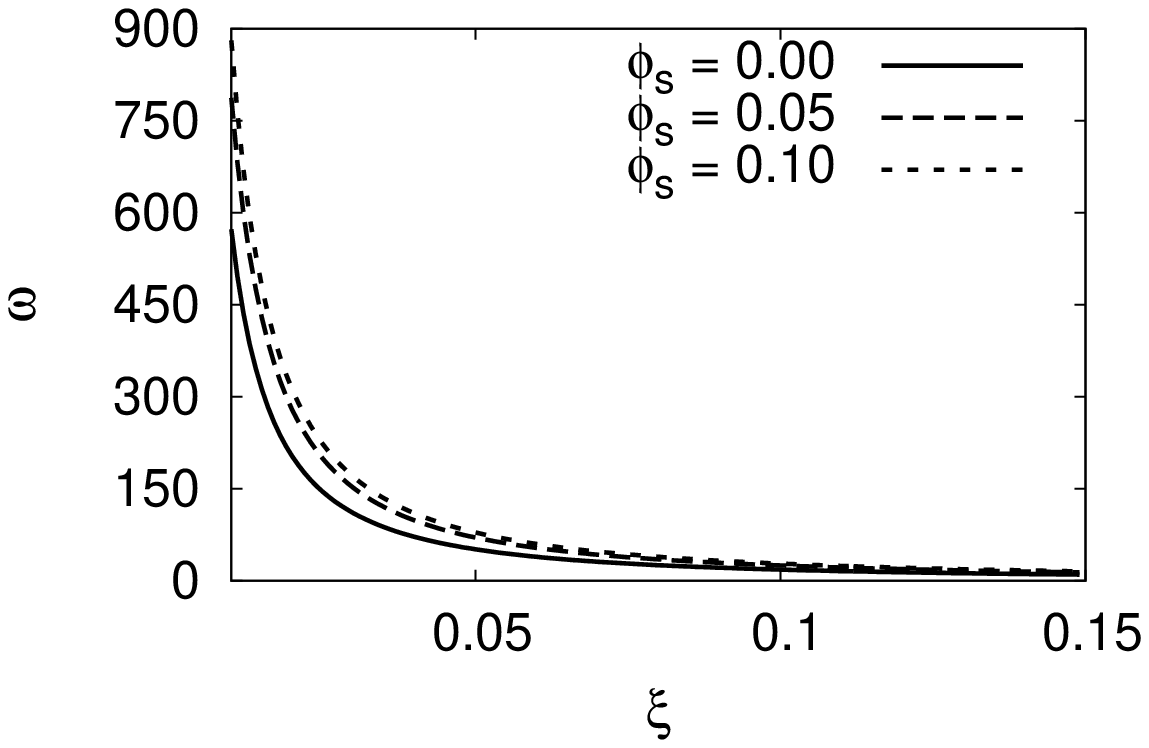}  }
}
  \caption{Time-dependent self-similar solution for $\gamma=1.3$, $\alpha=0.1$, $f=1.0$, and $\dot{m}_{in}=0.001$.
The solid, the dashed, and the short-dashed lines represent $\phi_s=0$, $0.05$,
and $0.1$, respectively.
}
\end{figure*}

\input{epsf}
\begin{figure*}
\centerline
{ 
{\epsfxsize=7.5cm\epsffile{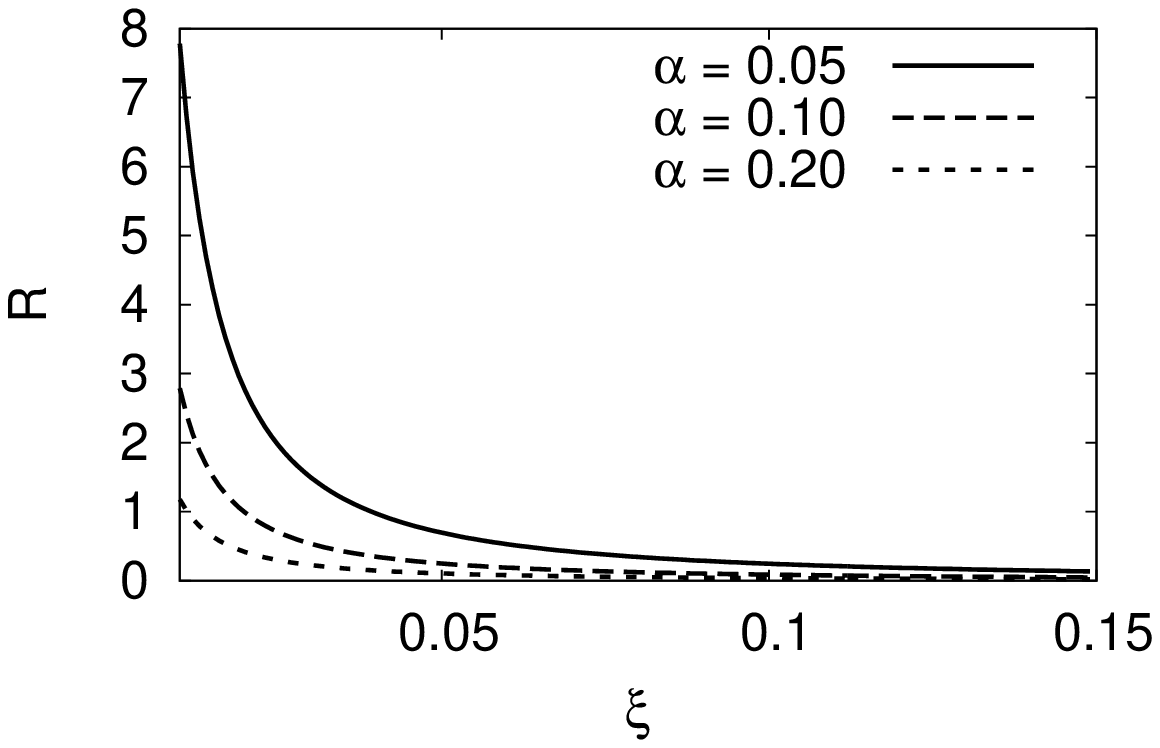}  }{\epsfxsize=7.5cm\epsffile{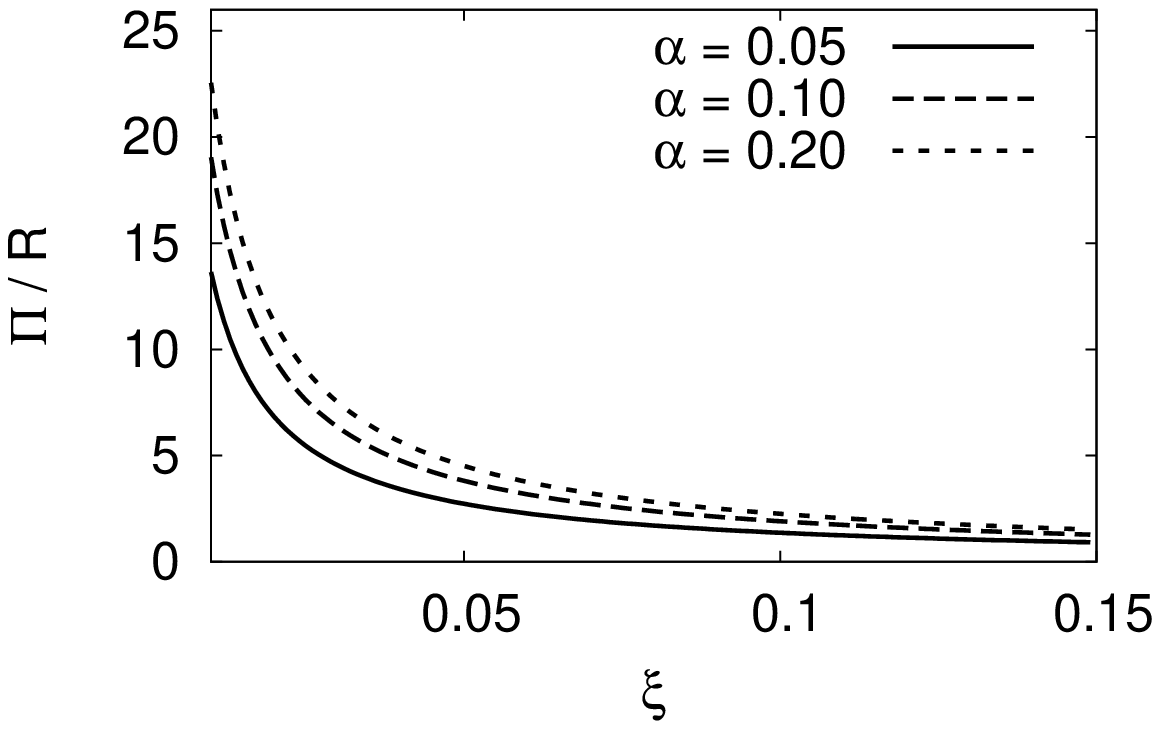}  }
} 
\centerline
{ 
{\epsfxsize=7.5cm\epsffile{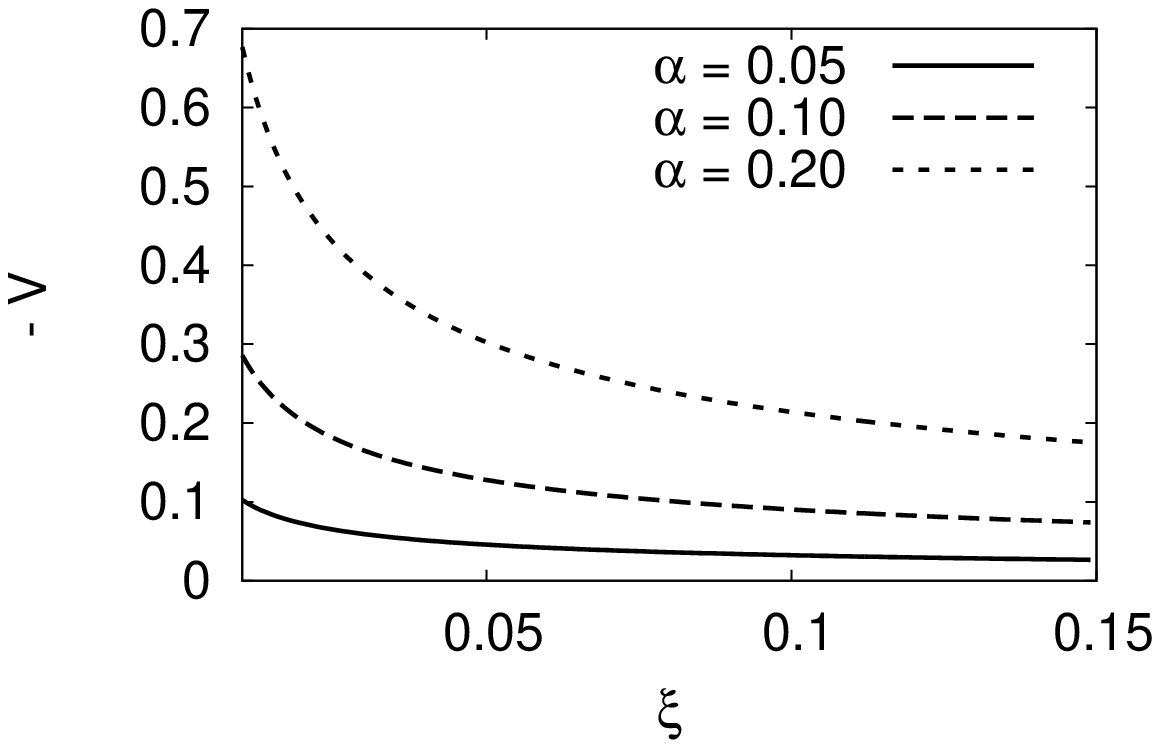}  }{\epsfxsize=7.5cm\epsffile{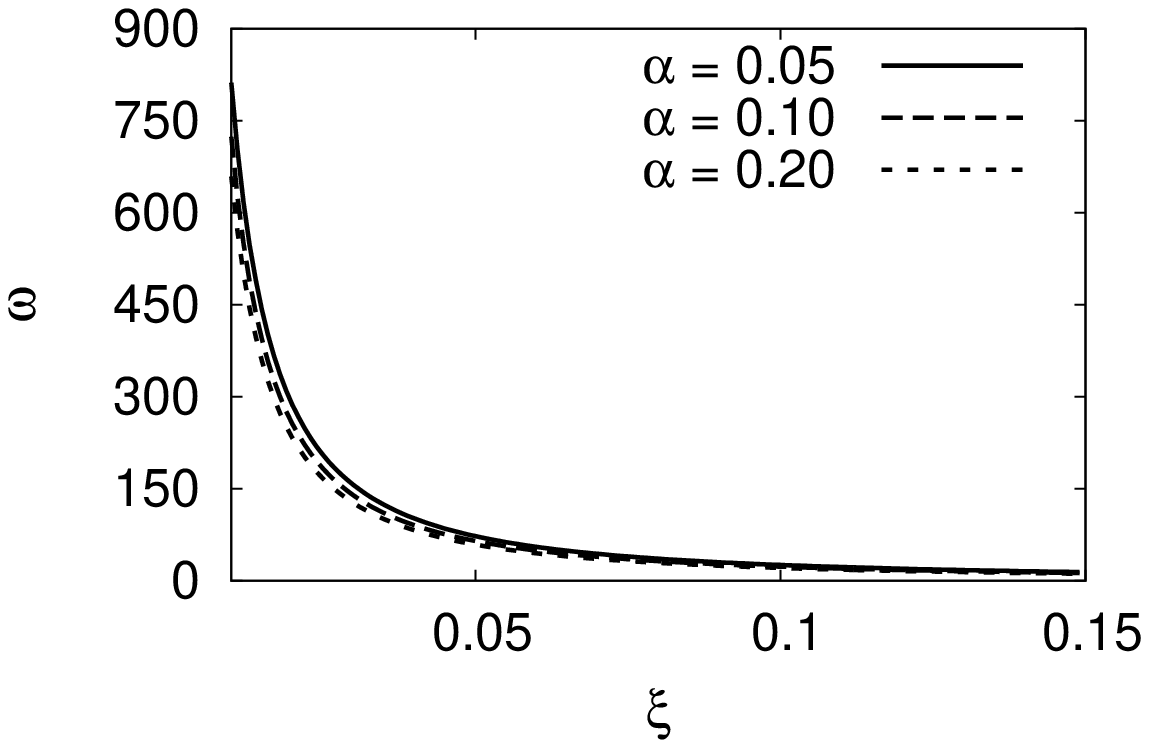}  }
}
  \caption{Same as Figure 1, but $\phi_s=0.03$.
The solid, dahsed, and short-dashed lines represent $\alpha=0.05$, $0.1$,
and $0.2$, respectively.} 
\end{figure*}

\subsection{Numerical solution}
If the value of $\xi_{in}$ is guessed, i.e. we take a point very near to the centre, the equations (23)-(26)
by Runge-Kutta-Fehlberg fourth-fifth order method
can be integrated  from this point outward by the above expansion [(27)-(30)]. Examples of such solutions
are presented in Figs 1-5. 

\subsubsection{The influences of saturation constant and viscous parameter on physical quantities}
The delineated quantity of $\Pi/R$ in Figs 1 and 2 is the sound speed square
in self-similar flow, which is rescaled in the course of time and represents
the flow temperature. The profiles of $\Pi/R$ in Fig. 1 show 
the flow temperature decreased by adding saturation constant, $\phi_s$.
Because, the generated heat by viscous dissipation can be transfered by thermal conduction.
Furthermore, this temperature decrease
is qualitatively consistent with simulation results of  Sharma et al. (2008)
and  Wu et al. (2010). 
We know the viscous dissipation of the flow increases by adding $\alpha$ parameter. 
Thus, the temperature increased by viscous parameter confirmed by the temperature profiles in Fig 2.
The density profiles show the gas density increased by adding the $\phi_s$ parameter.
It can be
due to temperature fall of fluid. Increasing density  by adding saturation constant is another consistency of our results with
simulations of Wu et al. (2010). Also, the density profiles show that it decreases by adding the viscous parameter.
This also could be result of the temperature rising. 
The viscous turbulence 
in this paper
is proportional to the gas temperature ($\nu \propto c_s^2 \propto T$). Thus,  
increase or decrease of temperature 
will affect dynamics of the accreting gas. As we said the temperature decreased by saturation constant
 implies that the viscous turbulence decreases, too. The decreasing of viscous turbulence reduces
the effect of negative viscous torque in
angular momentum equation. Thus, we expect the flow rotates faster by adding saturation constant that 
confirmed by the angular velocity profiles in Fig. 1. Since, the efficiency of the angular momentum transport decreases 
by adding the saturation constant, 
we expect the decrease of radial infall velocity 
that the radial velocity profiles in Fig. 1 confirm it.
The efficiency of angular momentum transport increases by adding the viscous parameter of $\alpha$. Thus, we expect
 the flow rotates slower and accretes faster by
adding the viscous parameter. The profiles of radial and angular velocities
in Fig. 2 confirm them.

\input{epsf}
\begin{figure*}
\centerline
{ 
{\epsfxsize=8.5cm\epsffile{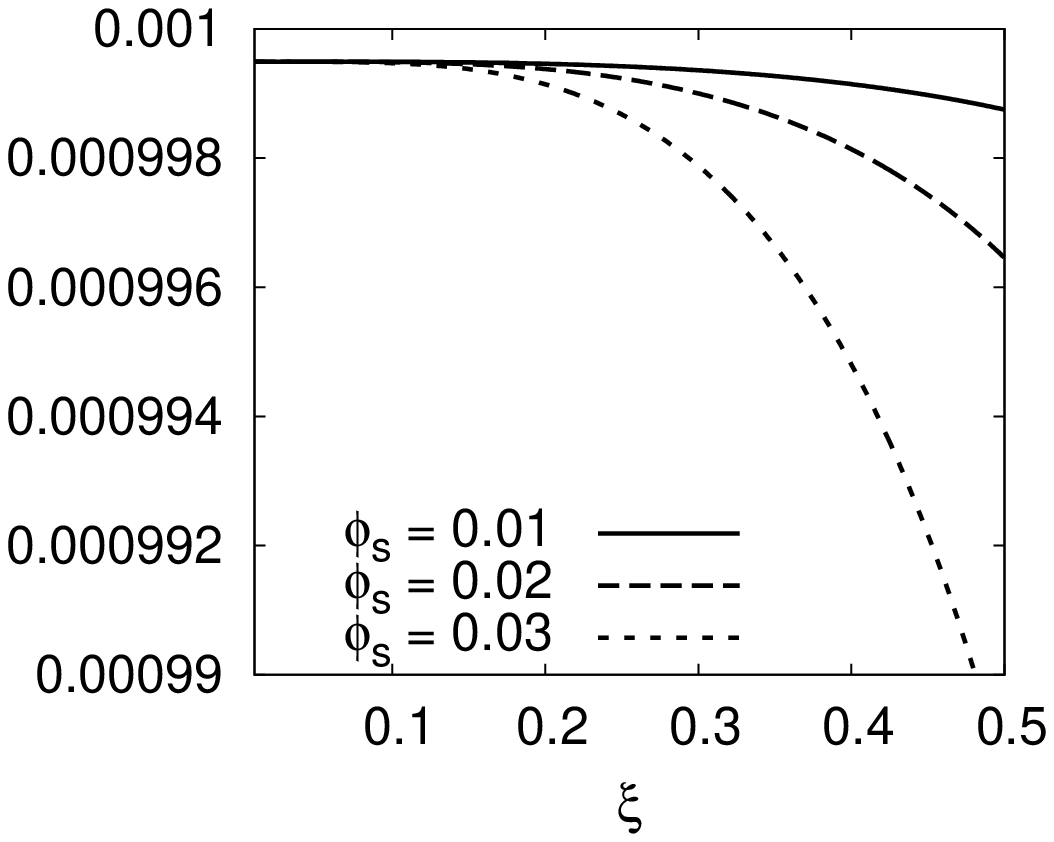}}{\epsfxsize=8.5cm\epsffile{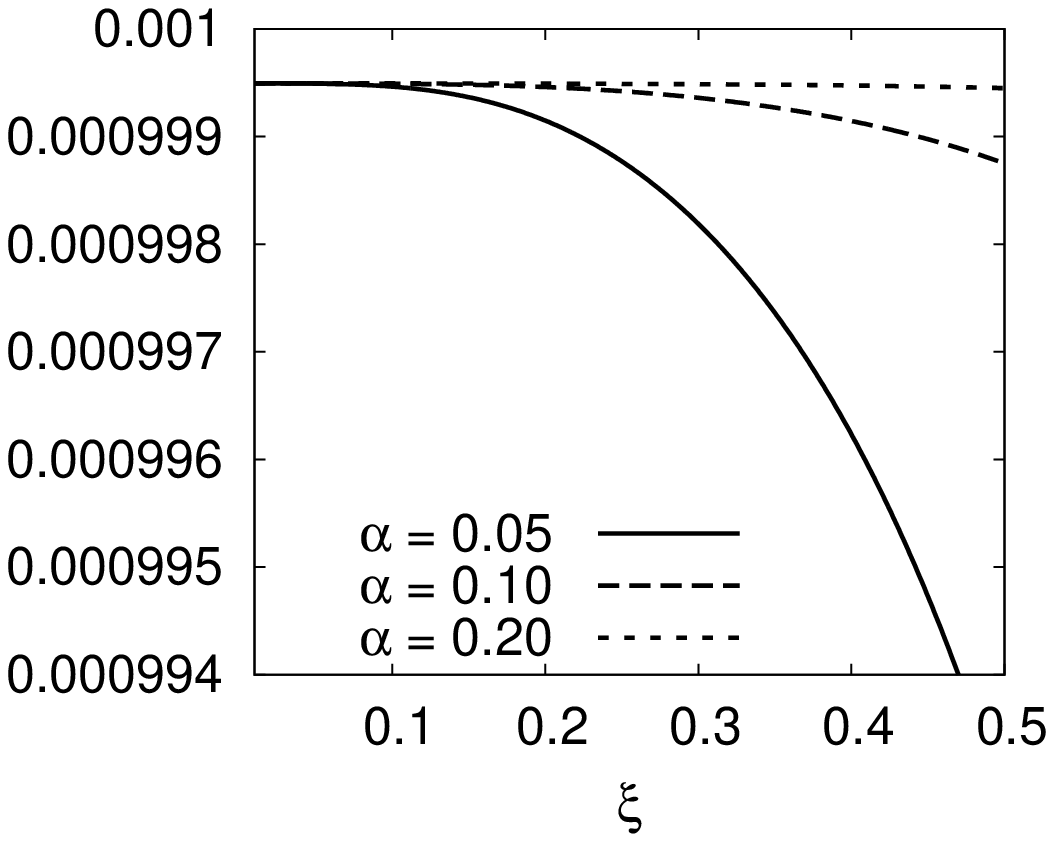}  }
} 
\caption{Time-dependent self-similar solution of mass accretion rate. The input parameters in
\textit{left panel} are same as Figure 1, but   
the solid, the dashed, and the short-dashed lines represent $\phi_s=0.01$, $0.02$, and $0.03$, respectively. The input parameters in
\textit{right panel} are same as Figure 2. but $\phi_s=0.01$.  
}
\end{figure*}

\subsubsection{Mass accretion rate}
The behaviour of mass accretion rate as a function of similarity variable $\xi$ 
for several values of the viscous parameter and 
saturation constant are
 plotted in Fig. 3. In the present model, 
the mass accretion rate 
is reduced by radius. While, the mass accretion rate in steady hot accretion flows is a constant (Tanaka \& Menou 2006). There are
some researches in steady hot accretion flows that have studied power-law function of mass accretion rate (Shadmehri 2008; Abbassi et al. 2008).
However, the mass accretion rate in their models is not dependent on important parameters such as saturation constant
 and viscous parameter. 
The profiles of mass accretion rate in Fig. 3 show that it is reduced by adding the saturation constant. This property is qualitatively 
consistent with numerical results of Johnson \& Quataert (2007).
Also, the profiles of mass accretion rate imply that it increases by adding the viscous parameter of $\alpha$.
This property is qualitatively consistent with previous works in accretion flows  
(e. g. Park 2009).

\input{epsf}
\begin{figure*}
\centerline
{ 
{\epsfxsize=8.5cm\epsffile{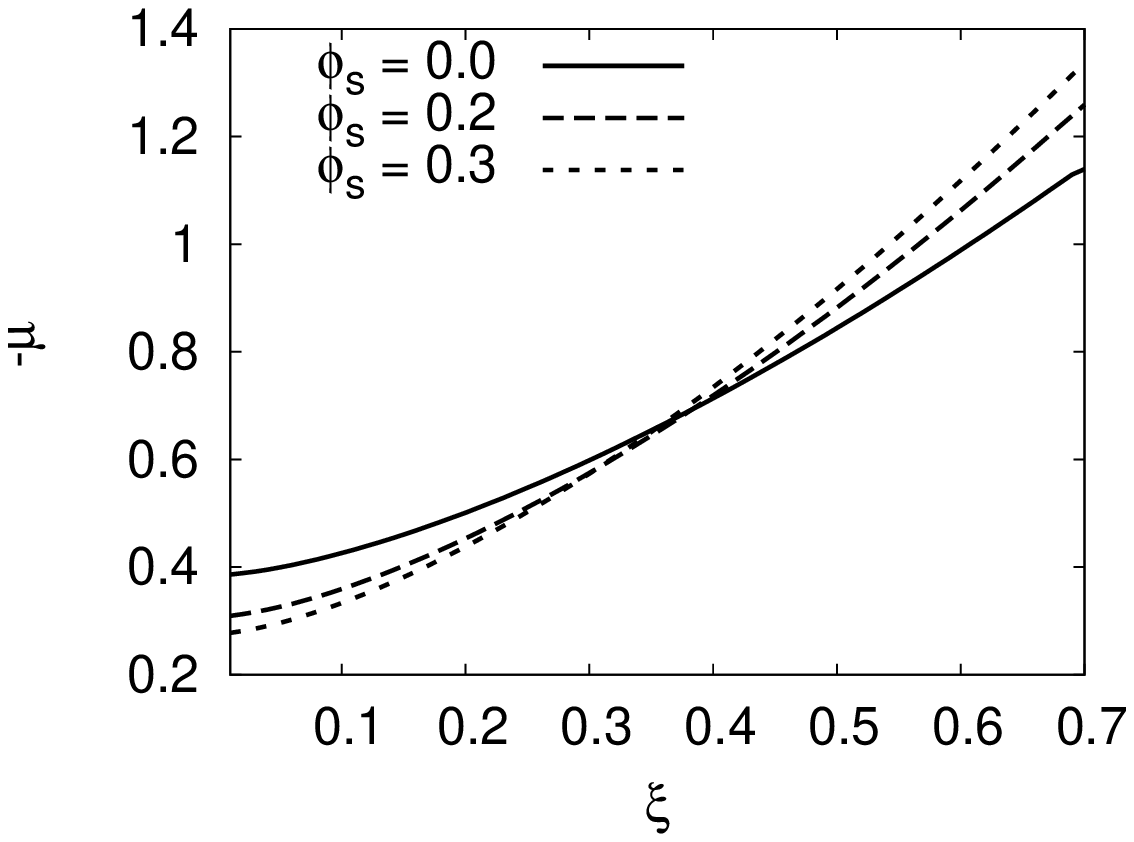}  }{\epsfxsize=8.5cm\epsffile{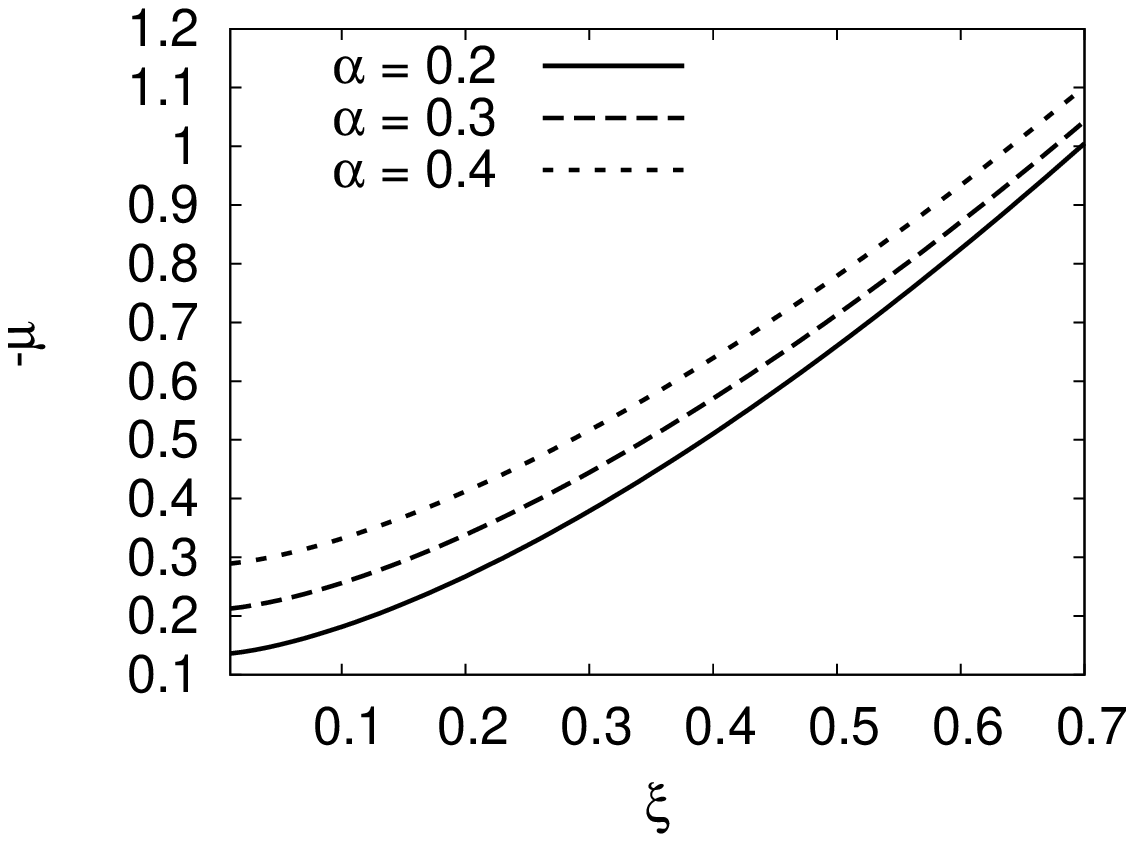}  }
} 
 \caption{Time-dependent self-similar solution of Mach number. The input parameters in
\textit{left panel} are same as Figure 1, but $\alpha=0.5$ and 
the solid, the dashed, and the short-dashed lines represent $\phi_s=0.0$, $0.2$, and $0.3$, respectively. The input parameters in
\textit{right panel} are same as Figure 1, but $\phi_s=0.05$ and  
the solid, the dashed, and the short-dashed lines represent $\alpha=0.2$, $0.3$, $0.4$, respectively. 
}
\end{figure*}

\subsubsection{Mach number}
Here, it will be interesting to investigate the existence of the transonic point in hot accretion flow.
The transonic point occurs in place that the amount of \textit{Mach number} becomes equal to unity.
The Mach number referring to
the reference frame is defined as (Gaffet \& Fukue 1983; Fukue 1984)

\begin{equation}\label{a36}
    \mu\equiv\frac{v_{r}-v_{F}}{c_{s}}=\frac{V-n\xi}{S}
\end{equation}
where
\begin{equation}\label{a37}
    v_{F}=\frac{dr}{dt}=n\frac{r}{t}
\end{equation}
is the velocity of the reference frame which is moving outward 
as time goes by, which the sound speed can be subsequently expressed as
\begin{equation}\label{a35}
c_{s}^2 \equiv \frac{p}{\rho}= S^2 (G M_*/ t)^{2/3}
\end{equation}
and, $S=\left(\Pi/R\right)^{1/2}$ the sound speed
in self-similar flow is rescaled in the course of time.
The Mach number introduced so far, represents the
\emph{instantaneous} and \emph{local} Mach number of the unsteady
self-similar flow.
 In steady self-similar solution (e. g. Tanaka \& Menou 2006),
the Mach number does not vary by radii and is a constant. While, the Mach number in
unsteady self-similar varies by radii (see Fig. 4).  As seen in Fig. 4, there is a transonic point ($|\mu|=1$).
The dependency of transonic point to
saturation constant 
shows that this point moves inward
by adding the parameter of $\phi_s$. 
Because, thermal conduction transfers the heat to larger radii, so 
the sound speed/temperature decreases by adding the saturation constant. In other words, 
whatever the radii smaller, the radial velocity relative to the sound speed larger.
 Also,
the Mach number profiles show the transonic point decreasing by adding the viscous parameter
which can be due to increase of radial velocity along with adding the viscous parameter.

\input{epsf}
\begin{figure*}
\centerline
{ 
{\epsfxsize=8.5cm\epsffile{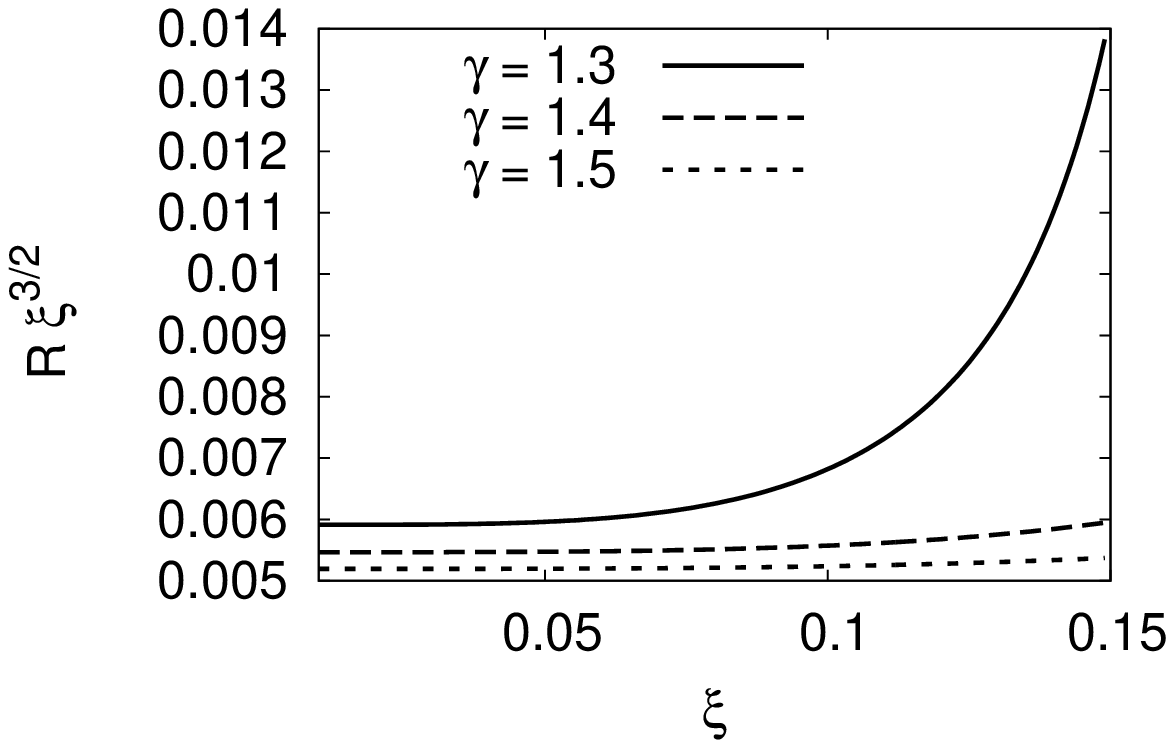}  }{\epsfxsize=8.5cm\epsffile{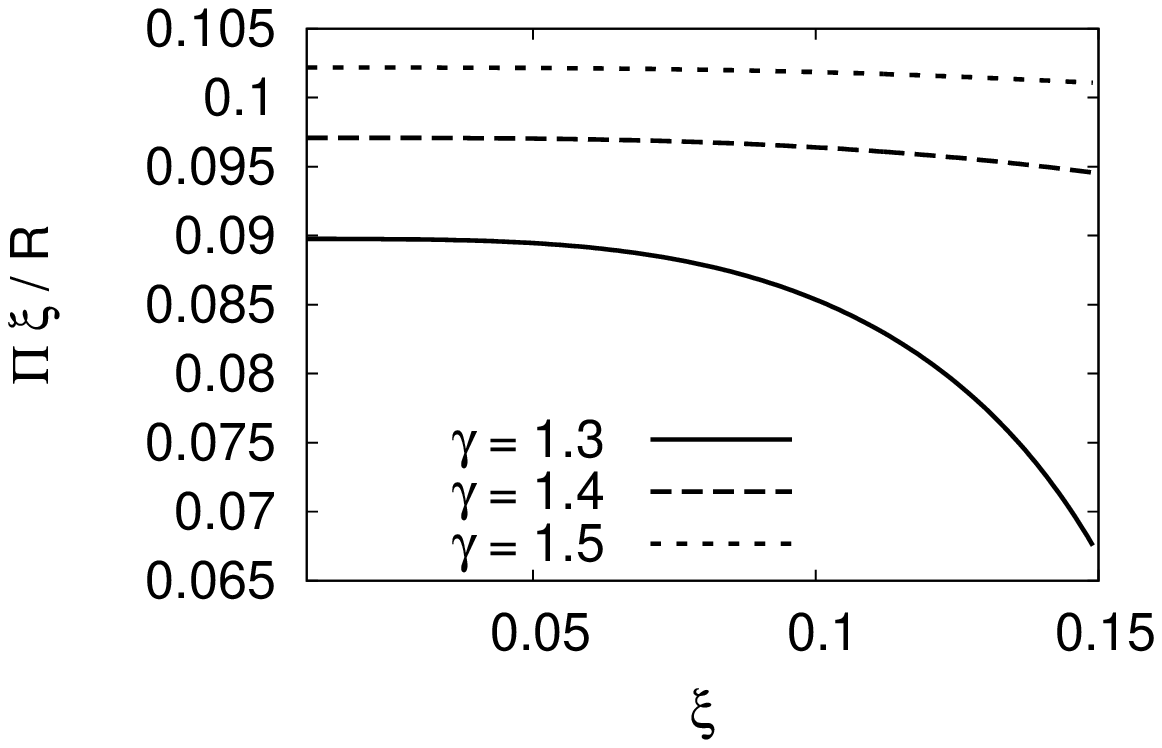}  }
} 
\centerline
{ 
{\epsfxsize=8.5cm\epsffile{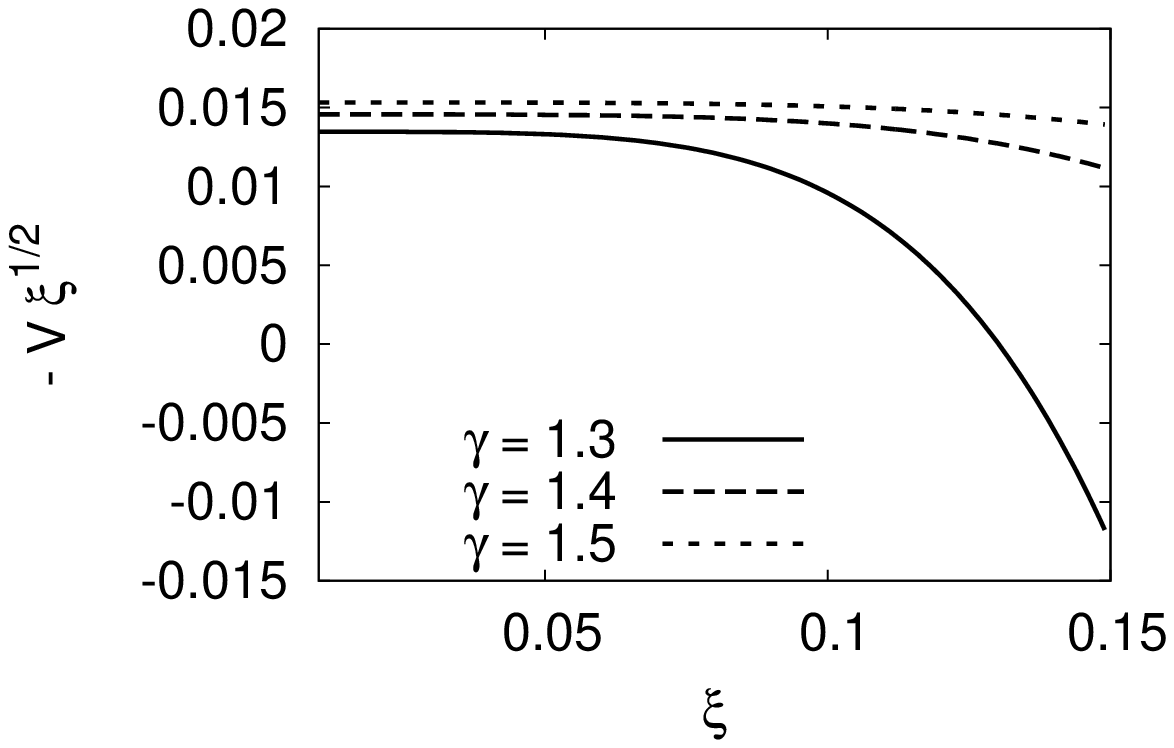}  }{\epsfxsize=8.5cm\epsffile{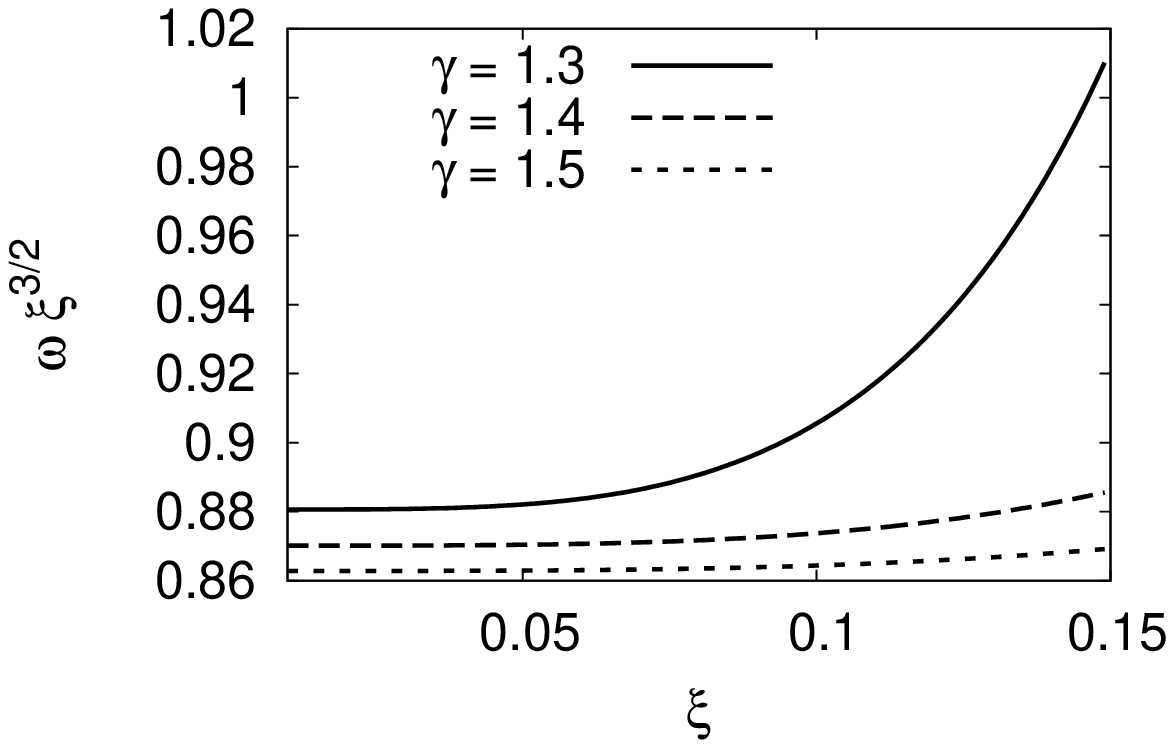}  }
}
  \caption{Same as Figure 1, but $\phi_s=0.1$.
The solid, the dashed, and the short-dashed lines represent $\gamma=1.3$, $1.4$,
and $1.5$, respectively.} 
\end{figure*}

\subsubsection{Comparison of steady and unsteady self-similar solutions}
As a comparison between steady and unsteady self-similar solution, the physical quantities 
in unsteady self-similar solutions
are divided into their radial dependence in steady self-similar solution then plotted 
in terms of $\xi$ in Fig. 5. Also, the effect of adiabatic index on hot accretion flow is investigated in Fig. 5.
The delineated quantities ($R\, \xi^{3/2}$,
$V\,\xi^{1/2}$, $\cdot$ $\cdot$ $\cdot$) in Fig 5 are constant in steady self-similar solutions of hot accretion flows 
(Tanaka \& Menou 2006; Shadmehri 2008; Abbassi et al. 2008; Ghanbari et al. 2009). While they
vary with position in this research. Fig. 5 represents the density profile varies shallower than $\xi^{-3/2}$, that this property
is qualitatively consistent with simulation results (e.g., Stone et al. 1999; Igumenshchev \& Abramowicz 1999;
Stone \& Pringle 2001; Hawley \& Balbus 2002; Igumenshchev
et al. 2003).  The radial dependency of the temperature and radial velocity in unsteady self-similar solution show 
that they vary deeper than the steady self-similar solution. Also, the study of the angular velocity show that   
it varies shallower than $\xi^{-3/2}$. Thus, 
the physical quantities in the present model avoid the limits
of the steady self-similar solution. The physical quantities profiles in Fig. 5 show
that their radial dependency in the unsteady self-similar limited to the steady self-similar solution by adding adiabatic index $\gamma$.

\section{Summary and Discussion}
In hot accretion flows, the collision timescale between ions and electrons is longer than the inflow timescale. Thus, the inflow
plasma is collisionless, and the transfer of energy by thermal conduction can be dynamically important. The low collisional rate of
the gas is confirmed by direct observation, particularly in the case of the Galactic centre (Quataert 2004; Tanaka \& Menou 2006)
and in the intracluster medium of galaxy clusters (Sarazin 1986).

Here, we have investigated how thermal conduction affects dynamics of hot quasi-spherical accretion 
flows. We adopted the presented solutions
by Ogilvie (1999) and Tanaka \& Menou (2006). Thus, we assumed that angular momentum transport is due to viscous turbulence 
and the $\alpha$-prescription 
is used for the kinematic  coefficient of viscosity. We also assumed the flow does not have a good cooling efficiency and so
a fraction of energy accretes along  with matter on to the central object. The effect of thermal conduction is studied by
a saturation form of it introduced by Cowie \& McKee (1977). To solve the equations that govern the dynamical behaviour
of hot accretion flow, we have used unsteady self-similar solution.   

The effect of saturation constant and the viscous parameter on the present model is investigated.
The solutions show that
with the increase of conductivity, the equatorial density becomes denser and the temperature becomes lower. These results
are qualitatively consistent with simulation results of Wu et al. (2010). Furthermore, the solutions show that
by adding the saturation constant, the angular velocity becomes larger and  the radial velocity decreases. 
The mass accretion rate is reduced by adding the saturation constant that is qualitatively consistent
with the result of Johnson \& Quataert (2007). The solutions imply that
the viscous parameter 
has opposite effects in comparison to saturation constant on physical quantities of the system. 
Also, the study of physical quantities of the present model in comparison to steady self-similar solution
show that our results deviate from steady self-similar solution and do not have its limits. 

Here, we studied dynamical behaviour of hot accretion flow in one-dimensional approach ignored by
latitudinal dependence of physical quantities.
Although, some authors have shown that latitudinal dependence of physical quantities is important for structure and dynamics of
hot accretion flow (Tanaka \& Menou 2006; Ghanbari et al. 2009; Wu et al. 2010). Thus, latitudinal behaviour of
the present model can be investigated in other studies.

\section*{Acknowledgments}
I would like to thank the anonymous referee for very useful comments that helped me to improve the initial version of the paper.



\end{document}